\documentclass[preprint]{aastex631}
\graphicspath{{./}{figures/}}

\usepackage{amsmath}
\usepackage{graphicx}
\usepackage{natbib}
\usepackage{breakurl}
\usepackage{multirow}
\usepackage{empheq}
\usepackage{float}
\usepackage{tabularx}
\usepackage{gensymb}

\usepackage{verbatim}  

\usepackage{CJK}

\usepackage{color}

\begin{document}
\begin{CJK*}{UTF8}{gbsn}

\title{Constraining Thermal Emission of Pluto's Haze From Infrared Rotational Lightcurves}


\author{Linfeng Wan (万霖丰)}
\affil{Department of Earth and Planetary Sciences, University of California Santa Cruz, CA 95064, USA}
\author{Xi Zhang}
\affil{Department of Earth and Planetary Sciences, University of California Santa Cruz, CA 95064, USA}
\author{Jason D. Hofgartner}
\affil{Southwest Research Institute, Boulder, CO 80302, USA}

\begin{abstract}

The rotational lightcurves of the Pluto-Charon system were previously believed to be solely attributed to their surfaces. However, a proposed scenario of haze cooling \citep{2017Natur.551..352Z} suggests that the atmospheric haze of Pluto could significantly contribute to mid-infrared emission, which calls for a revisit of previous analyses. In this study, we employ a Bayesian retrieval approach to constrain the haze emission from the rotational lightcurves of the Pluto-Charon system. The lightcurves were observed by the Spitzer and Herschel telescopes at 24 and 70 $\mu$m, and were combined with the latest surface albedo maps of Pluto and Charon from the New Horizons spacecraft. Our results show that including the haze emission is consistent with all current observations, with the best-fit haze flux around 1.63 mJy. This is in agreement with the composition of Titan-like tholins. However, the ``surface only" scenario, which excludes the haze contribution, can still explain the observations. We conclude that the current data at 24 $\mu$m cannot constrain Pluto's haze emission due to the degeneracy with Charon's surface emission. Regardless, some surface properties of Pluto are well constrained by the shape of the lightcurves, with a thermal inertia of approximately 8--10 MKS and a relatively low CH$_4$ emissivity of 0.3--0.5. We suggest that observations by the JWST telescope at 18 $\mu$m, which can resolve Pluto from Charon, could directly probe the haze emission of Pluto due to the low surface emission at that wavelength.

\keywords{planets: surface -- planets: Pluto -- techniques: retrieval}
\end{abstract}

\section{Introduction}
\label{sec:1}

Rotational lightcurves provide insights into the surface properties and inhomogeneities of Pluto and Charon as the surface units rotate in and out of view. While reflection lightcurves mainly reveal the body's albedo and scattering properties, emission lightcurves at infrared and radio wavelengths offer valuable information about the surface and subsurface, such as temperature, thermal inertia, and emissivity. Before the New Horizons flyby, Pluto's surface albedo distribution could only be resolved with poor precision using techniques such as Pluto-Charon mutual events \citep[e.g.,][]{1992Icar...97..211B, 1993Icar..102..134Y, 1999AJ....117.1063Y, 2001AJ....121..552Y} and direct imaging by HST \citep[e.g.,][]{1997Icar..125..233B, 1997AJ....113..827S, 2010AJ....139.1117B, 2010AJ....139.1128B}.

Nevertheless, by combining reflection and emission lightcurves from HST, ISO, Spitzer, and Herschel, \citet{2000Icar..147..220L, 2011Icar..214..701L, 2016A&A...588A...2L} found that Pluto's surface temperature is non-uniform. They also constrained the thermal inertia and emissivity of the surfaces of Pluto and Charon. Multiple solutions have been identified, primarily due to the inability to resolve the Pluto-Charon system and the various choices of Pluto's surface map (Figure 4 in \citealt{2011Icar..214..701L}). Although radio telescopes such as the Atacama Large Millimeter/submillimeter Array (ALMA), Submillimeter Array (SMA), and Karl G. Jansky Very Large Array (VLA) are able to observe Pluto and Charon individually, the seasonal effect and the mixing of subsurface and surface information at radio wavelengths have complicated data interpretation \citep{2016A&A...588A...2L}.

The New Horizons flyby of the Pluto-Charon system in 2015 provided a wealth of new information that calls for revisiting previous rotational lightcurve data. High-definition images of Pluto and Charon obtained by the Long Range Reconnaissance Imager \citep[LORRI,][]{2008SSRv..140..189C} and the Multispectral Visible Imaging Camera \citep[MVIC,][]{2008SSRv..140..129R} provided detailed spatial distribution of geological units, and further spectral analysis of data by the Linear Etalon Imaging Spectral Array \citep[LEISA,][]{2008SSRv..140..129R} revealed complex ice compositions on different surface units composed of N$_2$, CO, CH$_4$, H$_2$O, etc. It was found that Pluto's surface has complex volatile distributions that are closely related to ice sublimation, gas condensation, and glacial flow \citep{2016Sci...351.9189G, 2016Sci...351.1284M}. On the other hand,  Charon has a relatively uniform surface with a reddish northern polar region \citep{2016Sci...351.9189G, 2017Icar..287..124S}. The preliminary surface albedo maps of Pluto and Charon based on New Horizons data, derived by \citet{2017Icar..287..207B, 2019ApJ...874L...3B}, contain much greater detail than the surface maps in previous studies. Therefore, it is important to revisit previous thermal rotational lightcurve data based on the new albedo maps, which is the first motivation of this study.

The second motivation is related to the hazy atmosphere of Pluto. Traditionally, the rotational lightcurve analysis of the Pluto-Charon system was focused only on their surfaces. However, during the New Horizons flyby, it was discovered that Pluto's atmospheric temperature was much colder than previously thought based on the gas-only model. Proposed solutions include cooling from the high concentration of super-saturated water vapor \citep{2017Icar..291...55S} or from the atmospheric haze \citep{2017Natur.551..352Z, 2021ApJ...922..244W} that was clearly visible in the New Horizons images \citep{2016Sci...351.8866G}. If haze is the main coolant in Pluto's atmosphere, it is predicted that the thermal emission from the haze could significantly contribute to the observed mid-infrared flux, such as that observed at 23.68 (hereafter 24) $\mu$m by Spitzer \citep{2011Icar..214..701L}, while the haze contribution might be negligible in the far-infrared wavelengths like 71.42 (hereafter 70) $\mu$m by Spitzer \citep{2011Icar..214..701L} or longer wavelengths by Herschel \citep{2016A&A...588A...2L}. This effect is important if the haze particles behave optically similar to Titan-like tholins \citep{2017Natur.551..352Z}. Recent studies suggested that the haze particles might be primarily composed of hydrocarbon and nitrile ices, which might produce a much smaller cooling effect  \citep{2021NatAs...5..289L}. A later study found that the ice particles could also explain the cold temperature of Pluto's atmosphere \citep{2021ApJ...922..244W}. In this case, the infrared flux from the icy haze seems much smaller than the Titan-like organic haze, but its contribution to the thermal lightcurve might not be negligible. The large thermal emission from haze could obscure the surface features and greatly reduce the relative amplitude of rotational lightcurve variations in the mid-infrared. As a result, including the haze emission in the rotational lightcurve analysis would not only lead to a better understanding of how the haze affects the derivation of surface properties from observed lightcurves, but also provide an opportunity to constrain the haze emission, which is largely uncertain in atmospheric models due to the lack of constraints from composition and optical constants for icy particles (\citealt{2021NatAs...5..289L}, Section 2.2 and Discussion in \citealt{2021ApJ...922..244W}) and non-icy tholin-like materials \citep{1984Icar...60..127K, 2020Icar..34613774J, 2021Icar..36214398J, 2022JGRE..12706984M}.

To better constrain both the surface properties and haze emission, in this study we also introduce a Bayesian retrieval framework that can efficiently evaluate the posterior distribution of parameters. The Bayesian inference technique has been widely used in analyzing remote sensing data on solar system planets \citep[e.g.,][]{2013Icar..226..159Z, 2019E&SS....6.1057F} and exoplanets \citep[e.g.,][]{2011ApJ...738...32L, 2013ApJ...775..137L}. It has also been used to retrieve the chemical abundances in Pluto's atmosphere from occultation data \citep{2018Icar..300..174Y} as well as the haze distribution like particle sizes \citep{2022NatCo..13..240F}. However, it has not been applied to Pluto's rotational lightcurve analysis before. With detailed surface albedo maps of Pluto and Charon now available, using Bayesian inference might provide useful constraints on the posterior distribution of the derived surface properties and haze emission.

We analyze three rotational lightcurves of the Pluto-Charon system in the infrared, selecting data from Spitzer at 24 and 70 $\mu$m in 2004 \citep{2011Icar..214..701L}, as well as data from Herschel at 70 $\mu$m in 2012 \citep{2016A&A...588A...2L}. We select the 24 $\mu$m data as the haze emissions could be significant at this wavelength \citep{2017Natur.551..352Z}, while the 70 $\mu$m data was selected as haze emissions are not important at this wavelength (as explained in Section \ref{sec:2.3}). By using both channels, we wish to differentiate the contribution of haze in the lightcurves. We include the Herschel data because of its better quality compared to the Spitzer data. However, we do not include the Herschel data at sub-mm wavelengths \citep[such as 500 $\mu$m,][]{2016A&A...588A...2L} due to the significant subsurface contribution and seasonal effects that could complicate the haze flux retrieval analysis.

Our paper is organized as follows: In Section \ref{sec:2}, we introduce the surface maps of Pluto and Charon from New Horizons, our thermal physical model, as well as the retrieval models; In Section \ref{sec:3}, we discuss our retrieval results in two scenarios, with and without the haze contribution. We also use our best-fit models to predict the rotational lightcurves of several channels on JWST's Mid-Infrared Instrument (MIRI) in Section \ref{sec:3}; Finally, we conclude this paper in Section \ref{sec:4}.

\section{Methods}
\label{sec:2}

\subsection{Maps of Albedo and Emissivity}
\label{sec:2.1}

\begin{figure}[!htb]
    \centering
    \includegraphics[width=18cm]{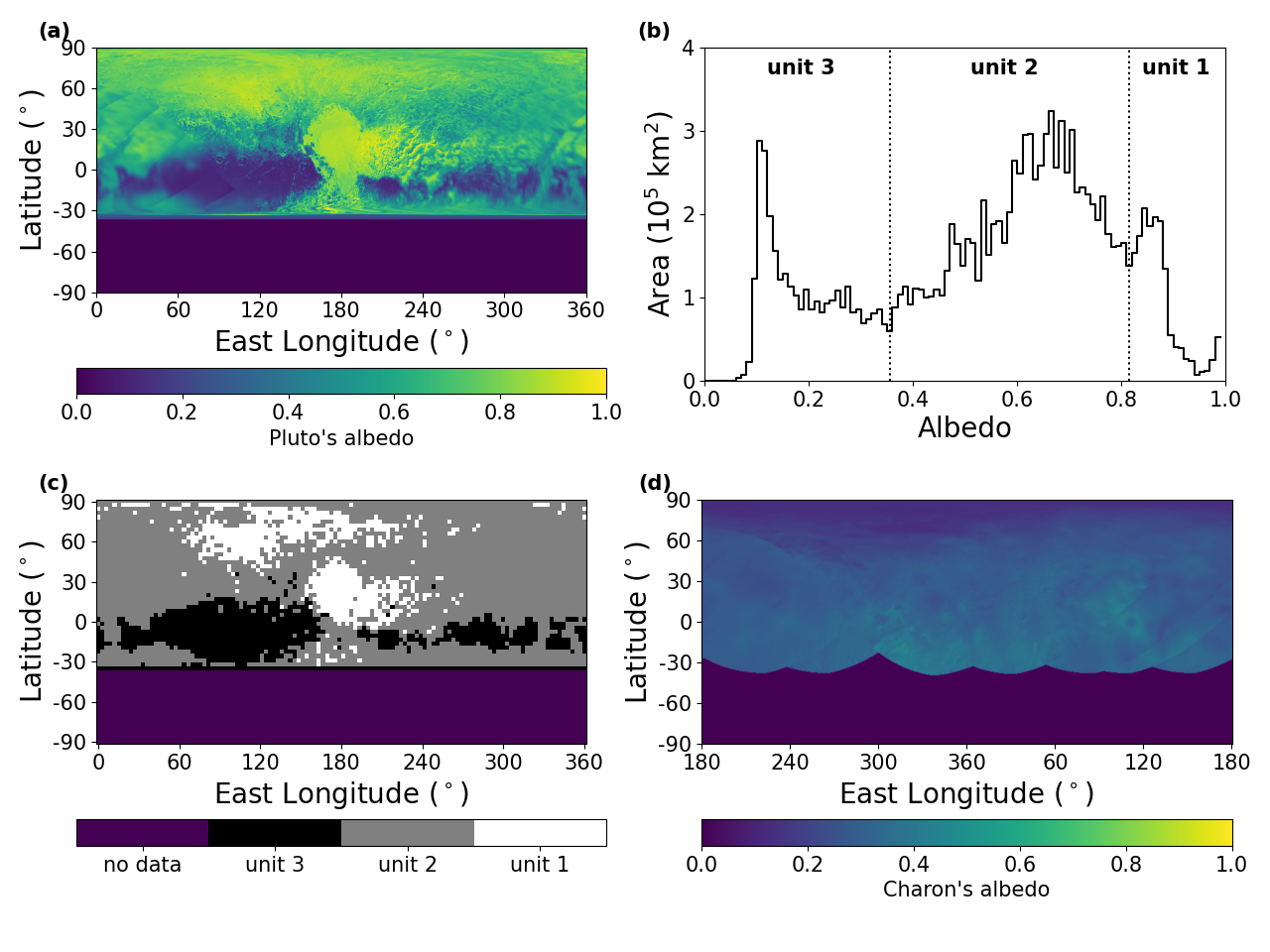}
    \caption{Surface maps of Pluto and Charon derived from New Horizons observations. Part of the southern hemispheres were in winter darkness (polar night) during the flyby, and thus the albedo distributions are not well determined south of $\sim$ -33$^\circ$. (a) Map of Pluto's bolometric hemispherical albedo from \citet{2023PSJ.....4..132H}. (b) Histogram of Pluto's albedo distribution, where local minima of 0.36 and 0.82 are adopted as unit boundaries. (c) The distribution of different albedo units on Pluto's surface. (d) Map of Charon's albedo from \citet{2017Icar..287..207B}. Note that, Charon rotates synchronously with Pluto and their lit hemispheres exhibit a 180$^\circ$ longitude difference.}
    \label{fig1}
\end{figure}

The surface temperatures of Pluto and Charon are greatly affected by albedo, emissivity, and thermal inertia. The distributions of all three parameters on Pluto are primarily determined by the distribution of volatile ices and non-volatile materials \citep{2016Sci...351.9189G, 2016Sci...351.1284M, 2017Icar..287..229S}. Surfaces of the Pluto-Charon system were observed during the New Horizons flyby in 2015, including high-resolution mosaics of the ``near side" and $\sim$20$\times$ poorer imaging of the ``far side" \citep{2017Icar..287..207B, 2021Icar..35613805S}. 

As a disk-integrated quantity, Bond albedo was widely used to measure the global energy balance of a planetary body. To calculate the local surface temperature, here we use the bolometric, hemispherical (disk-resolved) albedo. Preliminary maps of disk-resolved albedo for Pluto and Charon using panchromatic LORRI observations were presented in \citet{2017Icar..287..207B}. \citet{2023PSJ.....4..132H} updated the Pluto albedo map by fitting the lunar-Lambert photometric function \citep[e.g.,][]{1983Icar...55...93B} to both LORRI and MVIC measurements from a wide range of imaging geometries and by considering different photometric functions for Pluto's extreme dark and extreme bright terrains. The bolometric albedo was approximated using New Horizons LORRI panchromatic and MVIC panchromatic and color filter observations. The updated albedo map of Pluto is shown in Figure \ref{fig1}(a) and the map of Charon is shown in Figure \ref{fig1}(d).

In our study, we directly apply these latest albedo maps of Pluto and Charon from the New Horizons 2015 flyby to explain the 2004 Spitzer data and 2012 Herschel data. In other words, we assume the albedo distribution does not change between 2004 and 2015. For Pluto, this is only an approximation because the ice distribution on its surface might change with season \citep{2016Natur.540...86B, 2020JGRE..12506120B}. However, this is the only high-resolution map of Pluto's surface to date. Before the New Horizons flyby, Pluto's surface was barely resolved. Thus, \citet{2011Icar..214..701L} chose to use Pluto's HST reflection lightcurve in 2002--2003 to constrain the albedo of different geological units in their assumed surface maps. One might achieve a more accurate albedo map of Pluto in 2003 based on both the New Horizons and the HST data. That is beyond the scope of this study as it requires a detailed investigation using an atmosphere-surface model of Pluto \citep[e.g.,][]{2017Icar..284..443Y, 2020JGRE..12506120B}. As shown in Section \ref{sec:3}, using the New Horizons albedo maps can produce a decent fit of both the 2004 Spitzer and 2012 Herschel observations in our retrieval.

Due to the observing geometry, the New Horizons maps do not cover locations south of $\sim$ -33$^\circ$ for both Pluto and Charon. Because the sub-solar and sub-observer latitudes were 34.5$^\circ$ and 32.9$^\circ$ in 2004 (Spitzer) and about 47.0$^\circ$ in 2012 (Herschel), the missing albedos in the southern hemisphere would not strongly impact the lightcurves due to the projection effect. We tested a range of albedo values from 0 to 1 in those regions, and confirmed that our results were not changed. In our study, we utilize 120$\times$60 congrid maps for both Pluto and Charon, corresponding to a spatial resolution of 3$^\circ$. We also tested lightcurve simulations with higher spatial resolution, and the results were unchanged.

The emissivity distributions of Pluto and Charon are not directly constrained by New Horizons observations. For simplicity, we follow the approach in \citet{2011Icar..214..701L} that separated Pluto's surface into three geological units based on albedo and assumed a constant emissivity for each unit. We associate such composition units: dark H$_2$O ice/tholin mix (unit 3), CH$_4$ ice of intermediate albedo (unit 2), and bright N$_2$ ice (unit 1) with three albedo units in Figure \ref{fig1}(b) based on the histogram of Pluto's updated albedo distribution from New Horizons: unit 3 of darkest ices (0$-$0.36), unit 2 of intermediate ices (0.36$-$0.82), and unit 1 of brightest ices (0.82$-$1.0). The divided units (Figure \ref{fig1}c) appear to represent an approximate distribution of icy components identified by LEISA \citep{2016Sci...351.9189G, 2017Icar..287..229S, 2021Icar..35613833G, 2023PSJ.....4...15E}, although multiple ice mixtures show a much more complex pattern in the infrared LEISA maps. In contrast, Charon's surface (Figure \ref{fig1}d) is relatively uniform, and we use only one emissivity value. The thermal inertia distributions are also less constrained than that of albedo and we use a single value for Pluto and a separate single value for Charon.

\subsection{Thermophysical Model of Surface}
\label{sec:2.2}

Similar to previous work \citep{1989Icar...78..337S, 2000Icar..147..220L, 2011Icar..214..701L, 2017Icar..287...54F}, we model the surface and subsurface temperature distribution considering incoming solar heating, surface thermal emission, and thermal conduction for Pluto and Charon. Sublimation and deposition of volatiles are not included. At every point on the surface, the temperature ($T$) of subsurface layers at a time ($t$) and depth ($z$) obeys a diffusion equation:
\begin{equation}
\label{eqn:thermaldiffusion}
\rho c_p \frac{\partial T }{\partial t} =  \frac{\partial}{\partial z} ({ k \frac{\partial T}{\partial z}}) ,
\end{equation}
where $\rho$ (kg m$^{-3}$) is the density, $c_p$ (J K$^{-1}$ kg$^{-1}$) is the specific heat capacity and $k$ (W m$^{-1}$ K$^{-1}$) is the thermal conductivity. $\rho c_p$ = 10$^6$ J m$^{-3}$ K$^{-1}$ is adopted \citep{2017Icar..287...54F} everywhere at any depth on Pluto.
The upper ($z=0$) and lower ($z=\infty$) boundary conditions are 
\begin{empheq}[left=\empheqlbrace]{align}
& k \frac{\partial T}{\partial z} |_{z=0} +Q_s = E_s ,  \\    
& \frac{\partial T}{\partial z} |_{z=\infty} = 0 ,
\end{empheq}
respectively. At the surface, thermal conduction and the absorbed insolation ($Q_s$) balance with the infrared emission to space ($E_s$) \citep{2017JGRE..122.2371H}. We assumed no heat flux at the lower boundary.

To intuitively understand how deep the surface heat flux can propagate downwards to affect the subsurface temperature, we can introduce the skin depth ($l_{s}$)
\begin{equation}
\label{eqn:thermaldepth}    
l_s=\frac{\Gamma}{\rho c_p} \sqrt{ \frac{P}{\pi} } ,
\end{equation}
where $P$ is the period of stellar flux forcing and $\Gamma=\sqrt{k \rho c_p}$ is the thermal inertia. The depth of 8$l_s$ approximates the lower boundary, below which the periodic insolation changes hardly impact the substrate temperature \citep{1989Icar...78..337S}. 

Pluto has two dominant periods. For the diurnal cycle, $P_{day}$ = 6.4 Earth-days. If we assume the thermal inertia in the shallow subsurface is $\Gamma_{day}$ = 20 MKS (J s$^{-\frac{1}{2}}$ m$^{-2}$ K$^{-1}$) from \citet{2017Icar..287...54F}, the corresponding thermal diffusivity is $k$ = 4$\times$10$^{-4}$ W m$^{-1}$ K$^{-1}$ and the lower boundary depth is $8 l_s$ = 0.067 m. For the seasonal cycle, $P_{year}$ = 248 Earth-years. Thermal inertia of the deep subsurface $\Gamma_{year}$ = 800 MKS \citep{2017Icar..287...54F} corresponds to $k$ = 0.64 W m$^{-1}$ K$^{-1}$ and $8 l_s$ = 320 m, much deeper than the diurnal skin depth. Pluto is assumed to have homogeneous thermal inertia and conductivity; these values of $\Gamma$ and $k$ are used at all locations. Because our study focuses on rotational lightcurves, we only consider the shallow subsurface for the diurnal cycle in a self-rotating model, in line with \citet{1989Icar...78..337S}. In Appendix \ref{sec:A1}, we also test both diurnal and seasonal forcing using a two-layer, orbit-rotating model, which has a shallow subsurface with low thermal inertia overlaying a deep subsurface with high thermal inertia. The full-orbit simulations using this two-layer model show that the diurnal surface temperature variation in our self-rotating model is not greatly impacted by the seasonal cycle.

Evolution of the temperature is solved with a central difference explicit method and can be discretized by
\begin{equation}
T_j^{n+1} = T_j^n + dt \frac{k_j}{(\rho c_p)_j} \frac{1}{(z_{j+1}-z_{j-1})/2} ( \frac{T_{j+1}^n -T_j^n} {z_{j+1}-z_j} + \frac{T_j^n -T_{j-1}^n} {z_j-z_{j-1}} ) ,
\end{equation}
where the subscripts of $j$ and $n$ represent different levels and numerical time steps, respectively.
Similar to \citet{2017JGRE..122.2371H}, the boundary conditions with depth can be discretized by
\begin{empheq}[left=\empheqlbrace]{align}
& k_0 \frac{-3T_0^{n+1}+4T_1^{n+1}-T_2^{n+1}}{2(z_1-z_0)} = \epsilon \sigma {T_0^{n+1}}^4 -(1-A)F_0^{n+1} ,  \\
& T_N^{n+1} = T_{N-1}^{n+1} ,
\end{empheq}
where $A$ is the albedo (bolometric hemispherical albedo for Pluto and disk-resolved approximate Bond albedo for Charon), $\epsilon$ is the bolometric emissivity, $\sigma$ is the Stefan-Boltzmann constant, and $F_0$ (W m$^{-2}$) is the insolation that varies with time and location. For a given time and orbital phase, we first calculate the sub-solar point on Pluto using Equation (15) in \citet{2019ApJ...874....1O} and then the value of $F_0$ at any surface point based on its incident angle.

The model is divided into 15 uneven vertical layers, with the first two starting at a thickness of 10$^{-3}$ m, below which the thickness increases by a factor of 1.5 between adjacent layers. The total depth of the model is around 0.3 m, much deeper than $8 l_s$ (less than 0.1 m) for $\Gamma$ ranging from 20 to 30, as suggested in \citet{2011Icar..214..701L}. The thermal conduction timescale near the surface is $\sim$625 s, increasing with depth due to larger layer thickness. To accurately resolve a Pluto day during rotation, we set 1800 time steps with a time step of $\sim$300 s. The initial temperature profile is isothermal based on the surface's equilibrium temperature under diurnal-mean insolation. For each point on the surface, we run the model under diurnal insolation forcing until the diurnal variation of the surface temperature does not change between two subsequent rotations. Typically, five rotations are sufficient to reach convergence.

Given the albedo maps, emissivity, and thermal inertia, we can calculate the temporal and spatial distribution of the surface temperature of Pluto and Charon with our thermophysical model. However, the model does not consider the ice sublimation-condensation processes, which might impact the surface ice temperature through the latent heat exchange. Following \citet{2011Icar..214..701L}, we include additional treatment of N$_2$ and CH$_4$ ice temperature in our model. 
For the N$_2$ ice, given its high albedo (Figure \ref{fig1}) and bolometric emissivity of 0.5 \citep{2011Icar..214..701L}, the N$_2$ ice radiative-equilibrium temperature is low and is mainly controlled by the sublimation-condensation process. We can just fix the N$_2$ ice temperature to its vapor-ice equilibrium in our model. Based on the surface pressure of Pluto in 2004 and 2012 from stellar occultation and modeling \citep{2019A&A...625A..42M} and the saturation vapor pressure of the N$_2$ ice in the beta form \citep{2009P&SS...57.2053F}, we estimate the N$_2$ ice temperature to be 36.35 K in 2004 and 36.86 K in 2012.
For the CH$_4$ ice, it was found that the sublimation cooling effect becomes important when the CH$_4$ ice temperature exceeds $\sim$54 K \citep{1996P&SS...44.1051S}. Below 54 K, the radiative process controls the surface temperature. Thus, we set a maximum temperature of 54 K for the CH$_4$ ice, which was also adopted in \citet{2000Icar..147..220L, 2011Icar..214..701L}.

\subsection{Outgoing Thermal Emission}
\label{sec:2.3}

To model the rotational lightcurves, we first define the sub-observer point ($\alpha_0$, $\delta_0$), where $\delta_0$ is the latitude and $\alpha_0=\omega t$ is the longitude at time $t$ with a rotation rate $\omega=2\pi/P_{day}$. The outgoing surface emission ($F_{s}$) from Pluto and Charon at a given wavelength ($\lambda$), when viewed from the sub-observer point, can be calculated as follows:
\begin{equation}
F_{s,\lambda} (\alpha_0(t),\delta_0)=\int_{-\frac{\pi}{2}}^{\frac{\pi}{2}} \int_{0}^{2 \pi} \epsilon_{\lambda} (\alpha,\delta) B_{\lambda,T} (\alpha,\delta) \cos(\delta) \cos(\Theta) d\delta d\alpha,
\end{equation}
where $\epsilon_\lambda$ is the spectral thermal emissivity at wavelength $\lambda$, and $B$ (W m$^{-2}$ Sr$^{-1}$ $\mu$m$^{-1}$) is the blackbody emission determined by the surface temperature at a surface location ($\alpha$, $\delta$). $\cos(\Theta) = \max[\sin(\delta_0)\sin(\delta)+\cos(\delta_0)\cos(\delta)\cos(\alpha-\alpha_0), 0]$, where $\Theta$ means the angular distance between the sub-observer point ($\alpha_0$, $\delta_0$) and surface locations ($\alpha$, $\delta$).
We generate the rotational emission lightcurves by using a fixed sub-observer latitude ($\delta_0$) and varying the time $t$ from 0 to $2{\pi}/{\omega}$.

Previous thermal emission calculations only considered the contribution from Pluto's and Charon's surfaces \citep{2011Icar..214..701L} but neglected Pluto's haze flux, which might significantly affect the total outgoing emission. Pluto's haze extends several hundred kilometers above the surface, and the particles can scatter and absorb solar radiation in the optical wavelengths and emit in the mid-infrared. Due to the uncertainty in haze composition, the estimated haze infrared flux at 24 $\mu$m from atmospheric models differs by more than an order of magnitude, ranging from 0.1$\times$ to a few mJy 
 \citep{2017Natur.551..352Z,2021NatAs...5..289L,2021ApJ...922..244W}. The latter is comparable to the surface emission of Pluto at the same wavelength \citep{2011Icar..214..701L}. Thus, it is necessary to include Pluto's haze when analyzing the infrared spectra and thermal lightcurves of the Pluto-Charon system.

We need to consider several facts to account for the effects of Pluto's haze in the emission calculation. First, the haze might block part of the thermal emission from the surface, obscuring the surface features. However, the infrared optical depth of the haze is too small ($\sim10^{-3}$ at 24 $\mu$m, \citealt{2017Natur.551..352Z}) to be important. Therefore, we can directly add the haze emission to the surface emission in our model. 

Second, the haze emission contributes differently to the total outgoing flux at different wavelengths. The haze emission might be as important as the surface emission in the mid-infrared like 24 $\mu$m, but it is negligible in the far-infrared like 70 $\mu$m. The haze flux, calculated assuming Titan-like tholins in \citet{2017Natur.551..352Z}, is about 2 mJy at 70 $\mu$m, much smaller than the blackbody emission of Pluto's surface at this wavelength, which is over 100 mJy. As a result, in this study, we only consider the haze emission at shorter wavelengths such as 18 or 24 $\mu$m but not at longer wavelengths such as 70 $\mu$m.

Lastly, Pluto's global haze layers are not entirely spatially uniform \citep{2017Icar..290..112C} as shown in the north-south brightness asymmetry in the LORRI images. 3D haze transport simulation on Pluto \citep{2017Icar..287...72B} also shows that more haze was seen at the North pole due to the larger incoming solar flux. However, Pluto's gases and haze are predicted to be quickly mixed over all longitudes by the zonal circulation, therefore justifying the approximation of a longitudinally uniform haze layer. Because this study focuses on the disk-averaged flux, we assume a constant haze emission when Pluto is rotating.
The observed inverted temperature profile \citep{2017Icar..290...96H} above Pluto's cold icy surface was classically understood as being primarily caused by thermal conduction and CH$_4$ heating/cooling radiative equilibrium, with some additional role due to CO, HCN, and C$_2$H$_2$ cooling \citep{1989Natur.339..288Y, 1996Icar..120..266S, 2017Icar..291...55S}. However, it has recently been shown that haze radiative heating/cooling \citep{2017Natur.551..352Z} and eddy heat transport \citep{2021ApJ...922..244W} could dominate the region. If haze is the main coolant, the local surface temperature does not strongly influence the haze emission whose contribution mainly comes from 20 km above the surface where the atmospheric temperature maximizes \citep{2017Icar..290...96H}.

\subsection{Bayesian Inference Retrieval}
\label{sec:3.4}

Directly solving for the surface properties of Pluto and Charon, as well as Pluto's haze flux, is a challenging task due to the complexity of the thermophysical model and lightcurve calculations involved. We choose a Bayesian retrieval method, which has been widely used in planetary and exoplanet studies, to simultaneously constrain all parameters and assess their uncertainties. This method iteratively solves the inverse problem to find the best-fit model solution to the observed data. Various algorithms have been developed to fit observational data and determine the model parameters as well as their posterior distributions, such as Markov Chain Monte Carlo \citep{2003itil.book.....M} and Nested Sampling technique \citep{2004AIPC..735..395S}. For an efficient and unbiased sampling of the parameter space, we employ the well-developed software package \textit{PyMultiNest} \citep{2014A&A...564A.125B}, which utilizes the Nested Sampling technique. By comparing the simulated flux with observations from Spitzer and Herschel in \citet{2011Icar..214..701L,2016A&A...588A...2L}, we are able to find the best-fit model parameters that minimize the difference between the simulated and observed lightcurves. In addition, the Nested Sampling approach can estimate the posterior distribution (i.e., uncertainty) of the retrieved parameters.

We list important input parameters of our model in Table \ref{tab1}, including the distance and locations of the Sun and observer to Pluto in 2004 and 2012 \citep{2011Icar..214..701L, 2016A&A...588A...2L}, the radii of Pluto and Charon \citep{2017Icar..287...12N}, and the thermal properties of ices.
The retrieved parameters in our model include Pluto's thermal inertia ($\Gamma_{Pluto}$), emissivity of unit 2 (approximately CH$_4$ ice, $\epsilon_{CH_4}$), Charon's thermal inertia ($\Gamma_{Charon}$), as well as the haze flux at 24 $\mu$m ($F_{haze}$). We neglect the haze emission contribution at 70 $\mu$m as it is much smaller compared to the surface flux at that wavelength.
Because the N$_2$ ice temperature is low, its contribution to the rotational lightcurves is negligible. Thus, we fix its bolometric and spectral emissivity using the values in \citet{2011Icar..214..701L}. On the other hand, Pluto's H$_2$O ice/tholin mix unit (dark terrain) and Charon's surface generally have a high emissivity \citet{2011Icar..214..701L}, so we fix their emissivity as unity in our model.
We assume the spectral emissivity of the CH$_4$ ice at 24 and 70 $\mu$m is the same as the bolometric emissivity. In other words, there is only one emissivity value for the CH$_4$ ice. It appears that one emissivity parameter of the CH$_4$ ice unit in our retrieval model is sufficient to explain the observed lightcurves. Even though the CH$_4$ emissivity might be wavelength-dependent \citep{1996P&SS...44..945S} and previous studies \citep{2011Icar..214..701L} tested wavelength-dependent emissivity, the current data could not constrain all three independent values.

\begin{table*}[!htb]
\centering
\begin{tabular}{ |c||c|c|c| }
\hline
 & Spitzer (2004) & Herschel (2012) & JWST (2023) \\
\hline  \hline
Sub-solar distance (AU) & 30.847 & 32.19 & 34.6 \\
\hline
Sub-solar latitude ($^\circ$) & 34.5 & 47.0 & 58.3 \\
\hline
Sub-observer distance (AU) & 30.95 & 32.4 & 34.4 \\
\hline
Sub-observer latitude ($^\circ$) & 32.9 & 48.5 & 57.3 \\
\hline
Planetary radius (km) & \multicolumn{3}{c|}{1188 for Pluto and 606 for Charon} \\
\hline
Emissivity for N$_2$ ice & \multicolumn{3}{c|}{0.5 for bolometric and 70 $\mu$m, 0.05 for others$^{*}$} \\
\hline
Emissivity for H$_2$O ice/tholin mix & \multicolumn{3}{c|}{1.0 for bolometric and all wavelengths} \\
\hline
Emissivity for Charon & \multicolumn{3}{c|}{1.0 for bolometric and all wavelengths} \\
\hline
Fixed temperature for N$_2$ ice (K)$^{\dagger}$ & 36.35 & 36.86 & 36.35 \\
\hline
Max temperature for CH$_4$ ice (K) & \multicolumn{3}{c|}{54} \\
\hline
\end{tabular}
\caption{Input parameters of the thermophysical model. 
$^{*}$Based on the N$_2$ emissivities in \citet{2011Icar..214..701L}, we adopted 0.5 for bolometric emissivity and the emissivity at 70 $\mu$m, and 0.05 for other shorter wavelengths between 18 and 30 $\mu$m. $^{\dagger}$Due to the very low temperature and small spectral emissivity, this value does not affect Pluto's total thermal emission.}
\label{tab1}
\end{table*}

In the Nested Sampling approach, live points are the set of samples being actively updated by the algorithm during the course of the computation. At each iteration of sampling, the point with the lowest likelihood value (i.e., the least probable point) is removed from the set of live points and replaced with a new point that has a higher likelihood value. This process continues until the desired level of precision is achieved or until the algorithm converges. Increasing the number of live points can lead to more accurate estimates of parameters and their posterior distributions, but also increases the computational cost. The number of live points should be at least 10 times and preferably 25 times the number of free parameters. We choose 1600 live points for our retrieval, although 800 also works and gives similar results.

\begin{figure}[!htb]
\centering
\includegraphics[width=18cm]{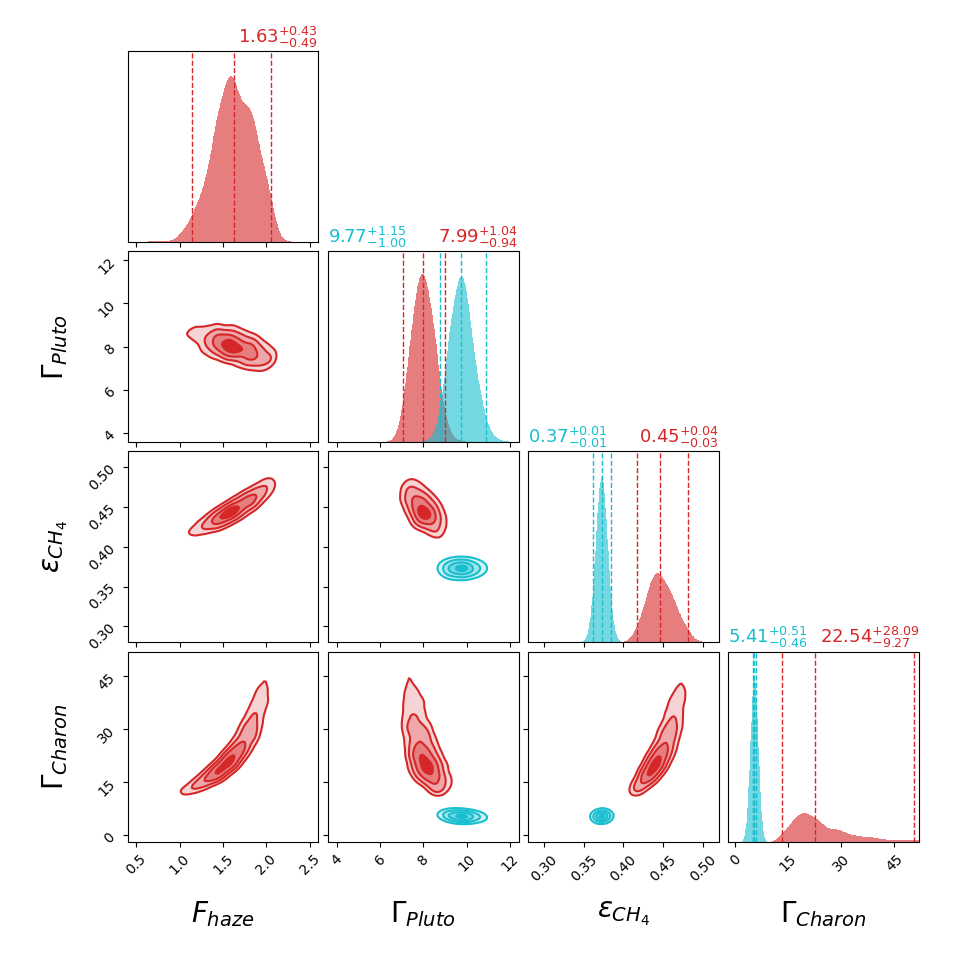}
\caption{The posterior distribution of retrieved parameters for the ``free retrieval" scenario with haze (red) and ``surface only" scenario without haze (cyan), including 24 $\mu$m haze flux $F_{haze}$ (mJy), CH$_4$ ice emissivity $\epsilon$, thermal inertia $\Gamma$ (MKS) for both Pluto and Charon. Vertical dashed lines in the diagonal panels denote 2.5\%, 50\%, and 97.5\% quantiles of the distributions. Off-diagonal panels show the two-dimensional parameter covariance plots, indicating a strong negative correlation between the haze flux and Charon's thermal inertia (lower-left panel).} 
\label{fig2}
\end{figure}

\begin{figure}[!htb]
\centering
\includegraphics[width=18cm]{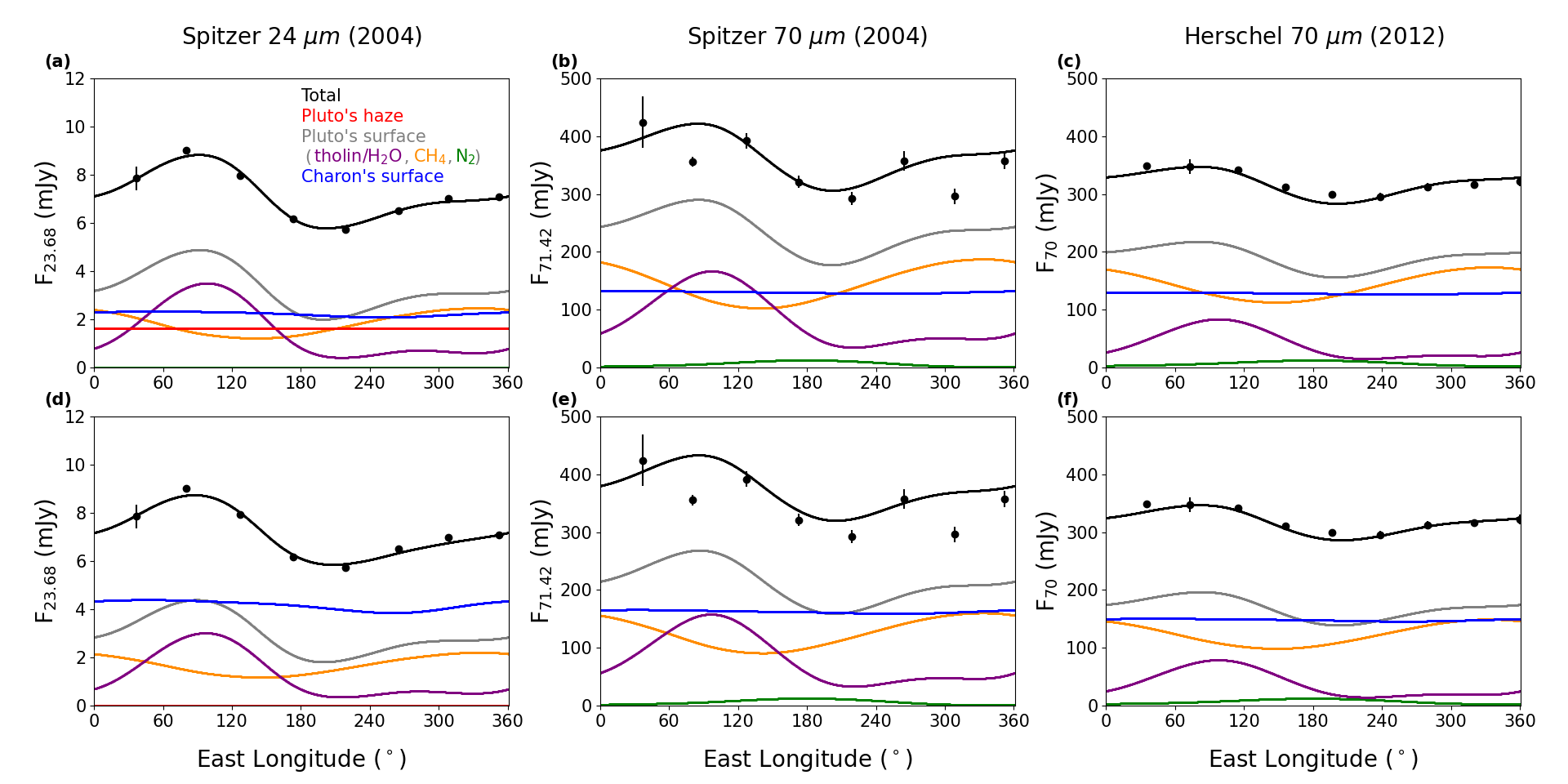}
\caption{Rotational emission lightcurves of Pluto and Charon from our best-fit models (lines) and observations (dots) at Spitzer 24 $\mu$m and 70 $\mu$m (first two columns), Herschel 70 $\mu$m (third column). Contributions from Pluto's different components are separately plotted for the ``free retrieval" scenario with haze (upper panel) and ``surface only" scenario (lower panel).}
\label{fig3}
\end{figure}

\begin{figure}[!htb]
\centering
\includegraphics[width=18cm]{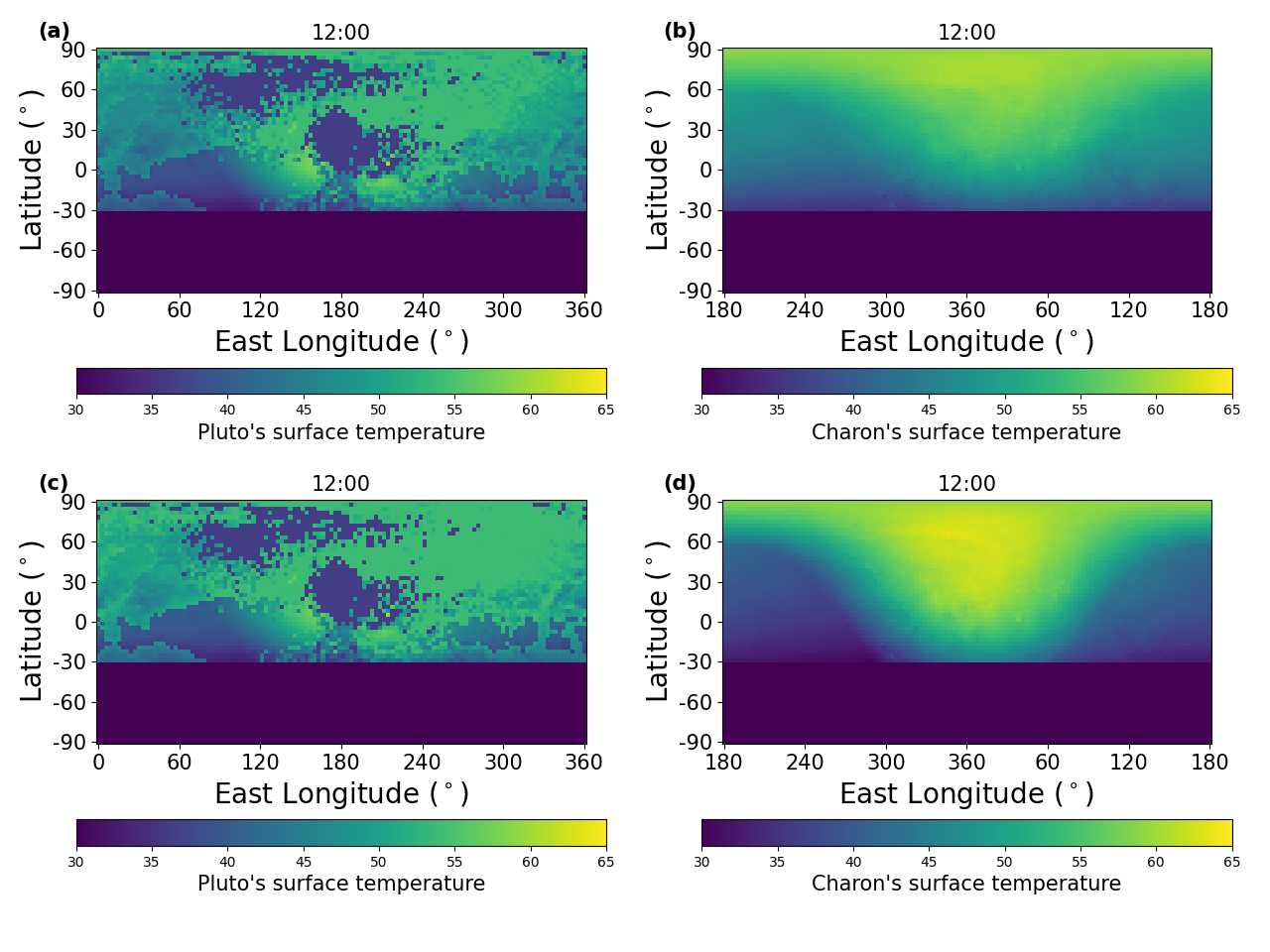}
\caption{Surface temperature distributions of Pluto (left) and Charon (right) in 2004 from our best-fit models in the ``free retrieval" scenario with haze (upper panel) and the ``surface only" scenario (lower panel). We plotted the maps at noon time when the sub-solar point at Pluto is 180$^\circ$ East Longitude. The animations of diurnal temperature evolution in 2004 and 2012 are available on the journal's website.}
\label{fig4}
\end{figure}

\section{Results}
\label{sec:3}

\subsection{Two Possible Scenarios to Explain Lightcurves}
\label{sec:3.1}

We first present the results from the ``free" retrieval, where we include the haze flux as a free parameter to constrain its value based on the observed lightcurves. The resulting posterior distribution of retrieved parameters and the fitted rotational lightcurves are presented in Figure \ref{fig2} (red) and Figure \ref{fig3} (upper panel), respectively.

Using the best-fit parameters, our model is able to explain the rotational lightcurves relatively well at both 24 and 70 $\mu$m. Both Pluto and Charon emit significantly more at 70 $\mu$m than at 24 $\mu$m, as shown in Figure \ref{fig3}. The corresponding temperature exhibits significant diurnal and spatial variation. We show the snapshots of their temperature maps at the noon time in Figure \ref{fig4}. We also provide animations of the diurnally-varying temperature distributions of Pluto and Charon from our best-fit models, which are available on the journal's website.

The lightcurves (black in Figure \ref{fig3}) at 24 and 70 $\mu$m exhibit two peaks, with the main peak occurring near 100$^\circ$ and a smaller peak near 300$^\circ$ in longitude. The main peak is primarily determined by the distribution of the dark and warm H$_2$O ice/tholin mix (purple) on Pluto, which is unevenly distributed along the longitude at low latitude regions (unit 3 in Figure \ref{fig1}). One can also see the corresponding temperature map at noon time when during Spitzer's observations in 2004 (Figure \ref{fig4}a). While the CH$_4$ ice, which covers a larger surface area in the northern latitudes (unit 2 in Figure \ref{fig1}), emits comparable flux to that of the H$_2$O ice/tholin mix but shows less variation across longitude. This CH$_4$ ice unit contributes to the second but smaller emission peak near 300$^\circ$ in longitude. In contrast, the N$_2$ ice emission (green) is negligible due to its much lower temperature. As the lightcurve variation primarily arises from Pluto's surface, it provides strong constraints on parameters such as Pluto's thermal inertia and emissivity. Unlike Pluto, Charon's uniform but darker surface (blue) produces almost flat lightcurves of about 2.2 mJy at 24 $\mu$m and 128.5 mJy at 70 $\mu$m, which is smaller than Pluto's total surface emission (gray).

In the best-fit model, Pluto's haze that was ignored in previous work \citep{2000Icar..147..220L, 2011Icar..214..701L, 2016A&A...588A...2L}, emits comparably to Charon's emission at 24 $\mu$m.
The haze emission of 1.63 mJy at 24 $\mu$m is similar to that in the atmospheric model calculations using Titan-like tholins in \citet{2017Natur.551..352Z}. However, the posterior distribution of this parameter indicates that the retrieved haze flux is 1.63$^{+0.43}_{-0.49}$ mJy, given by live points distributed within the 2.5\% and 97.5\% quantiles of the posterior by the Nesting Sampling technique (Figure \ref{fig2}). The lower value of 1.14 mJy at 24 $\mu$m might correspond to the situation of less absorbing particles in \citet{2017Natur.551..352Z}. Note that, in our model the haze contribution is negligible at 70 $\mu$m (see Section \ref{sec:2.3}).

The wide range of retrieved haze flux on Pluto is due to the inherent degeneracy between Pluto's haze emission and Charon's surface emission. This is because Spitzer and Herschel telescopes cannot resolve Charon from Pluto and both Pluto's haze and Charon's surface are spatially uniform. Even though we have tried to include the haze emission at 24 $\mu$m but not at 70 $\mu$m, and Charon contributes to both wavelengths, the current data is insufficient to break the degeneracy. As illustrated in the corner plot that shows the joint distribution of the parameters (lower-left panel of Figure \ref{fig2}), Pluto's haze emission exhibits a strong negative correlation with Charon's thermal inertia. For instance, when the haze flux is approximately 1.14 mJy at 24 $\mu$m, the retrieved Charon's thermal inertia is about 13 MKS, yielding about 2.87 mJy. If the haze flux is as high as 2.06 mJy, the corresponding derived Charon's thermal inertia is 45 MKS, corresponding to a Charon surface flux of about 1.75 mJy. This correlation implies that the total emission from Pluto's haze and Charon's surface is roughly constant in all solutions, which is about 4 mJy at 24 $\mu$m (Figure \ref{fig3}).


Moreover, in any retrieval process, the goodness of fit is influenced not only by data quality but also by the accuracy of the forward model. Our thermophysical model has made several assumptions, including dividing Pluto's surface into three geological units and a single thermal inertia value for the entire surface. Due to these model assumptions and incomplete information regarding the input albedo distribution (and neglecting its temporal evolution), our best-fit model does not perfectly match all data points in the observations, particularly those with minimal error bars (e.g., Figures \ref{fig3}b and \ref{fig3}e). Recognizing the potential biases in the forward model's construction and the substantial uncertainty range in the retrieved haze emission flux, we should be careful when interpreting the posterior distribution of the retrieved parameters. As an extreme scenario, it would be interesting to see the best fit using a model without haze. Therefore, we conduct an additional retrieval in a ``surface-only" scenario, where the haze flux is excluded, and only surface properties are retrieved. This approach resembles the one in \citet{2011Icar..214..701L}, albeit with updated albedo maps from New Horizons.

The parameters and their posterior distributions from the ``surface only" retrieval are shown in Figure \ref{fig2} (cyan). The fitted rotational lightcurves are shown in Figure \ref{fig3} (lower panel). Interestingly, our results are consistent with that in \citet{2011Icar..214..701L}. The ``surface only" scenario using the New Horizons albedo maps can also fit the rotational lightcurves relatively well without any haze contribution. As expected, the ``surface only" retrieval yields low thermal inertia of Charon (5.5$\pm$0.5 MKS), much lower than the ``free retrieval" results and previous estimates when using a different albedo map of Pluto \citep{2011Icar..214..701L, 2016A&A...588A...2L}. It is still acceptable because trans-Neptunian objects (TNOs) have even smaller values of 2.5$\pm$0.5 MKS \citep{2013A&A...557A..60L}. The contribution of Charon's emission flux at 24 $\mu$m is about 4.2 mJy in the ``surface only" case, roughly equal to the total emission of both Pluto's haze and Charon's surface in the ``free retrieval" case. Given that the ``free retrieval" scenario prefers a non-zero haze flux and that the ``surface only" scenario can also provide a decent fit, we conclude that the current data cannot constrain Pluto's haze flux. It also means that the current data cannot distinguish whether Pluto's haze is Titan-like tholins \citep{2017Natur.551..352Z} or ice particles \citep{2021NatAs...5..289L, 2021ApJ...922..244W}.

Radio observations of the Pluto-Charon system have a high spatial resolution that allows for the separation of Charon from Pluto. For example, SMA observed the dayside brightness temperature of Charon at 1.36 mm, yielding a temperature of 43$\pm$13 K in 2005. Our best-fit models give Charon's surface brightness temperature of 51.1 K for the ``free retrieval" scenario with haze and 54.0 K for the ``surface only" scenario without haze, respectively. Both temperatures are within the uncertain range of the SMA results. ALMA observations at 0.86 mm showed a temperature of 43.9$\pm$0.3 K for Charon in 2015, whereas our models predict a temperature of 51.8 K for the ``free retrieval" scenario with haze and 54.3 K for the ``surface only" scenario without haze.
These derived temperatures are greater in the scenario absent of haze due to smaller thermal inertia from retrievals, however, temperatures from both scenarios are larger than the corresponding radio observations. Similar results are also found for other observations, such as SMA at 1.10 mm in 2010 and VLA at 9.00 mm in 2011 (see Table 10 in \citealt{2019Icar..322..192B}), New Horizons dayside-staring at 4.17 cm in 2015 \citep{2019Icar..322..192B}, and Herschel at 500 $\mu$m in 2012 \citep{2016A&A...588A...2L}. This discrepancy is expected because radio observations do not directly probe the surface temperature but rather a mix of contributions from the surface and subsurface, which could be colder than the surface on the dayside. Furthermore, the radio emissivity of surfaces is not unity \citep{2016A&A...588A...2L}. Therefore, current radio observations of Charon's surface cannot distinguish the ``free retrieval" scenario with haze versus the ``surface only" scenario.

Observations of near-infrared spectra can also be used to probe the surface ice temperature, such as using the absorption bands of H$_2$O ice as thermometers \citep{1999Icar..142..536G, 2021psnh.book..433P}. \citet{2000Icar..148..324B} inferred a disk-average temperature of 60$\pm$20 K for Charon's surface via this method. Recently, \citet{2017Icar..284..394H} revisited this effort on the 1.65 $\mu$m absorption band of H$_2$O ice and reported Charon's mean surface temperature as 45$\pm$14 K in 2015. Our modeled surface temperature for both scenarios (49.5 K with haze and 48.9 K without haze in 2012) falls within this large uncertainty range, implying that the current near-infrared ``thermometer" is still too large to precisely constrain Charon's surface temperature to separate the two scenarios in this study.

While we cannot well constrain the properties of Pluto's haze and Charon's surface, we can determine their total emission flux at 24 $\mu$m in 2004, which is about 4 mJy in both haze and haze-free scenarios. In other words, the emission from Pluto's surface can be constrained. Indeed, Pluto's surface properties appear well-determined from the shape of lightcurves. The surface temperature distributions of Pluto in the best-fit models appear very similar between the ``free retrieval" scenario with haze (Figure \ref{fig4}a) and the ``surface only" scenario without haze (Figure \ref{fig4}c).
In both scenarios, the retrieved bolometric emissivity of Pluto's CH$_4$ ice is about 0.4, suggesting that the CH$_4$ ice surface might have small grains according to Figure 5 in \citet{1996P&SS...44..945S}. Our retrieved CH$_4$ emissivity is smaller than that from \citet{2011Icar..214..701L, 2016A&A...588A...2L}, which yields the bolometric emissivity of 0.7--0.9, spectral emissivity of 1.0 at 24 $\mu$m and 0.9 at 70 $\mu$m. The derived thermal inertia of Pluto is 10--20 MKS in \citet{2011Icar..214..701L} and 16--26 MKS in \citet{2016A&A...588A...2L}, higher than our optimal value (8--10 MKS). The reason has yet to be explored but is likely related to the difference in albedo maps in the two studies.

In summary, incorporating haze flux into retrievals is consistent with all existing observations. However, the current data quality is insufficient to distinguish between the contributions of Charon and haze, and cannot directly constrain the haze emission of Pluto. In the next section, we suggest that future observations using the JWST at additional wavelengths may assist in resolving this issue.

\begin{figure}[!htb]
\centering
\includegraphics[width=18cm]{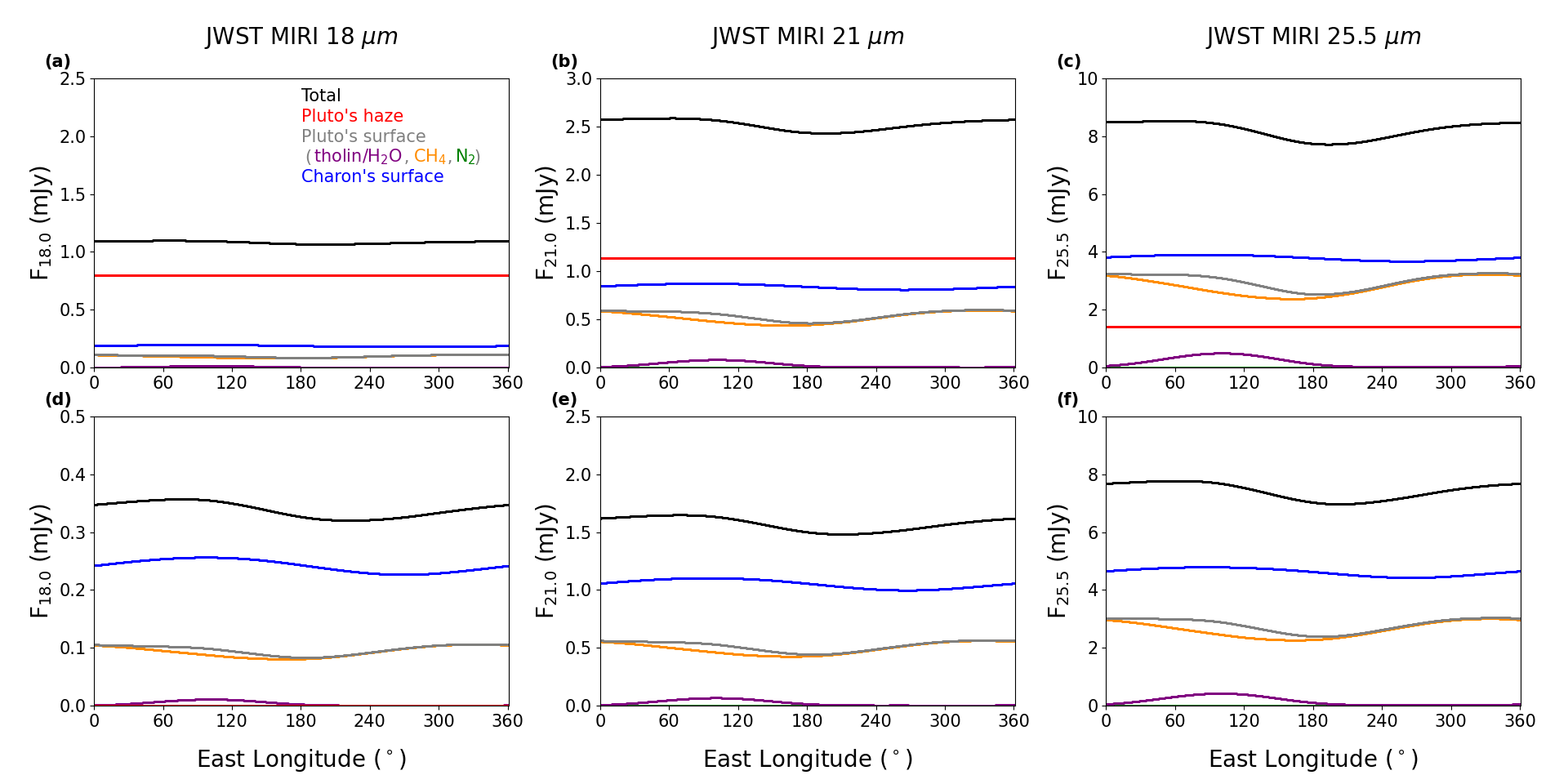}
\caption{Rotational emission lightcurves of Pluto and Charon simulated for the JWST observations at 18 $\mu$m, 21 $\mu$m, and 25.5 $\mu$m. Results of the ``free retrieval" scenario with haze (upper panel) and the ``surface only" scenario without haze (lower panel) are plotted with contributions from different geologic units and haze on Pluto and the surface of Charon. The haze emission fluxes at multiple wavelengths are calculated using the atmospheric model from \cite{2017Natur.551..352Z} assuming Titan-like tholins.}
\label{fig5}
\end{figure}

\begin{figure}[!htb]
\centering
\includegraphics[width=18cm]{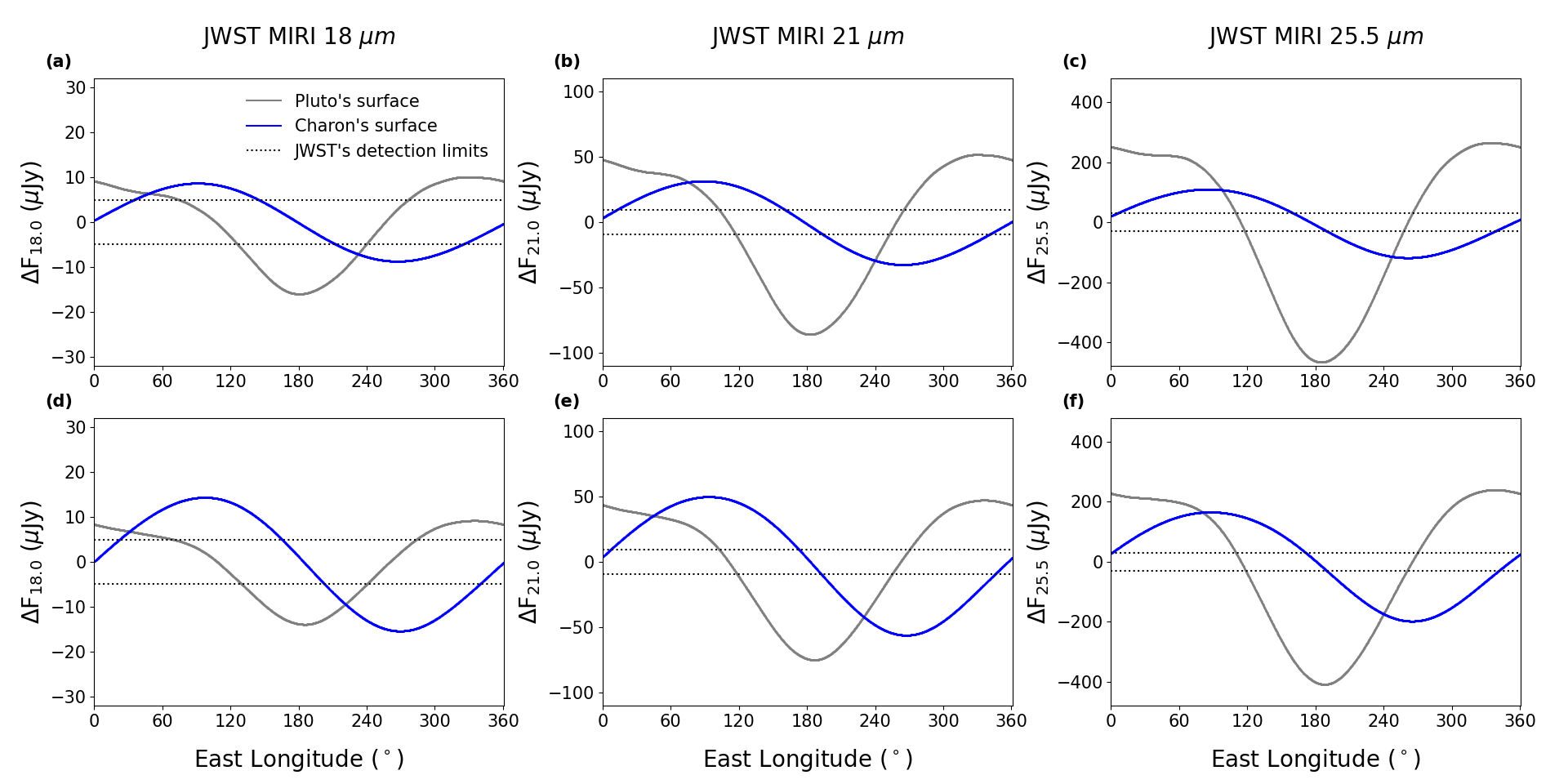}
\caption{Variation of rotational emission lightcurves (in units of $\mu$Jy) simulated for JWST's observations at 18 $\mu$m, 21 $\mu$m, and 25.5 $\mu$m. Results of the ``free retrieval" scenario with haze (upper panel) and the ``surface only" scenario without haze (lower panel) are plotted. Dashed lines indicate the 10$\sigma$ faint-source detection limits in a 10$^4$-second integration for corresponding channels of the JWST/MIRI.}
\label{fig6}
\end{figure}

\subsection{Predictions for JWST}
\label{sec:3.2}

Future observations with JWST can help distinguish between the two scenarios and provide valuable information on the haze and surface properties of Pluto. To achieve this, we have selected three typical channels of the JWST/MIRI: 18, 21, and 25.5 $\mu$m, with sensitivities \citep{2015PASP..127..686G} of 4.87, 9.15, and 30.8 $\mu$Jy, respectively. These channels are chosen as we primarily focus on the thermal emission from the Pluto-Charon system, and shorter wavelengths may be contaminated by reflected light. To support the planned JWST observations of Pluto in 2022--2023 \footnote{https://www.stsci.edu/jwst/science-execution/program-information.html?id=1658}, we calculate the rotational lightcurves based on a sub-solar latitude of 58.3$^\circ$, a sub-observer latitude of 57.3$^\circ$, a distance of 34.6 AU to the Sun, and a distance of 34.4 AU to the observer of JWST. In this configuration, it is possible to resolve Charon from Pluto and break the degeneracy between Pluto's haze emission and Charon's surface emission. Two limiting cases without haze and with thick haze (assuming Titan-like tholins) are presented in Figure \ref{fig5} using the best-fit parameters in the ``surface only" and ``free retrieval" scenarios in Section \ref{sec:3.1}, respectively. We assume the spectral emissivity of the surface units at these three wavelengths is the same as that at 24 $\mu$m from the retrieval.

In the ``surface only" scenario (lower panel in Figure \ref{fig5}), the surface emission from both Pluto and Charon increases with wavelength as it approaches the blackbody peak. At 18 $\mu$m, Pluto's surface emission is 0.1 mJy, which increases to 0.5 mJy at 21 $\mu$m and to 3.0 mJy at 25.5 $\mu$m. Similarly, Charon's surface emission increases from 0.25 mJy at 18 $\mu$m to 1.0 mJy at 21 $\mu$m and to around 4.5 mJy at 25.5 $\mu$m. Notably, Charon's surface emission is approximately twice larger than that of Pluto's surface for all three wavelengths. The contributions of different surface ices to Pluto's surface emission look dramatically different from those in 2004. This is because the sub-solar and sub-observer latitudes have moved much further north, from $\sim$34$^\circ$ in 2004 to $\sim$58$^\circ$ in 2023. Since most of the H$_2$O ice/tholin mix is concentrated near the equator and Pluto's northern hemisphere is predominantly covered with CH$_4$ ice, the emission of Pluto's surface (gray) mainly comes from the CH$_4$ ice (orange) rather than the H$_2$O ice/tholin mix (purple) in 2023 (see Figure \ref{fig5}). Therefore, the total lightcurves (black) are less variable compared to that in 2004 and 2012 (Figure \ref{fig3}) and peak at 300$^\circ$ in 2023 rather than 100$^\circ$ in longitude.

Taking Pluto's haze into consideration would make it more complicated to interpret the data, but it also offers an opportunity to constrain Pluto's haze emission. If the haze is made of absorbing Titan-like tholins, calculations using atmospheric model \citep{2017Natur.551..352Z} predict that it could emit much more than Pluto's surface at 18 and 21 $\mu$m (upper panel in Figure \ref{fig5}). JWST observations at these two wavelengths can be used to directly constrain the haze emission. On the other hand, if the haze emission is smaller, the surface emission and haze emission cannot be easily disentangled. As an extreme example, if the haze is made of brighter ice particles as suggested by the chemical models \citep{2021NatAs...5..289L}, the haze radiative cooling could be ten times smaller than that from the Titan-like tholins \citep{2021ApJ...922..244W}. For instance, if the tholin haze opacity is reduced by 20 times in the lower atmosphere as suggested by \citet{2021ApJ...922..244W}, such haze emission in the mid-infrared is comparable to the surface emission at 18 $\mu$m (0.1 mJy), a few times smaller than the surface at 21 $\mu$m, but negligible compared with the surface at 25.5 $\mu$m. Thus it is best to use 18 $\mu$m to constrain the haze properties. Yet, the emission from icy haze could also be large within its solid-state absorption band, for instance, within the 10-15 micron range for nitrile ices (see Figure 4 in \citealt{2017Natur.551..352Z}). Future spectra from the JWST may be capable of identifying these spectral structures. In any case, the rotational lightcurves from JWST in 2023 appear much flatter than the Spitzer observations as sub-observer latitude moves further north in absence of contribution from the H$_2$O ice/tholin mix. It might be difficult to distinguish the CH$_4$ ice emission at high latitudes and the haze emission. Combining with previous data from Spitzer or Herschel might be important to break the degeneracy, although assumptions need to be made such as that the ice distributions and haze temperature do not change significantly with time from 2004 to 2023. 

Interestingly, although the lightcurves of both Pluto and Charon exhibit nearly flat profiles in 2023, their variations can be characterized using JWST. In contrast to the lightcurves observed in 2004 and 2012, where Pluto's variation was much greater than that of Charon (Figure \ref{fig3}), our predictions indicate that Pluto's lightcurve variation during the JWST era will be comparable to that of Charon (Figure \ref{fig6}). At wavelengths of 18, 21, and 25.5 $\mu$m, the peak-to-trough flux amplitudes are approximately 20, 100, and 400 $\mu$Jy, respectively (Figure \ref{fig6}), all of which exceed the JWST/MIRI instrument's detection limit. Pluto's lightcurve exhibits a distinct ``U" shape, primarily influenced by the distribution of ice concentrated near the 180$^\circ$ mark in the northern hemisphere. On the other hand, Charon's lightcurve displays a maximum-flux phase shift from Pluto's, likely influenced by slightly darker materials in its northern high latitudes (Figure \ref{fig1}d) and the thermal phase lag (note that both Pluto and Charon are retrograde rotators with an obliquity larger than 90 degrees). Although distinguishing between best-fit models with and without haze in the lightcurve variation may be challenging (Figure \ref{fig6}), the shape of the lightcurve can serve as a useful constraint for determining the thermal inertia and ice albedo distributions on both bodies during the JWST era.

\section{Summary and Discussions}
\label{sec:4}

In this paper, we investigate the properties of Pluto's surface and haze using thermophysical modeling and Bayesian retrieval methods. We re-analyze the rotational lightcurves of the Pluto-Charon system observed by Spitzer and Herschel, incorporating Pluto's haze emission and the latest surface albedo maps from New Horizons. Our main findings are as follows:

1. Including Pluto's haze emission is consistent with all current observations. The best-fit model from our retrieval with haze yields about 1.63 mJy haze flux at 24 $\mu$m. In this case, Pluto's haze emission could be significant in the mid-infrared (e.g., 18--25.5 $\mu$m) compared with the surface emission of Pluto or Charon. 

2. However, the haze flux has a large uncertainty due to the degeneracy of 24 $\mu$m thermal emission between Pluto's haze and Charon's surface. Even without haze emission, the rotational lightcurves can be explained by the surface emission from Pluto and a low-thermal-inertia Charon. Therefore, the current data cannot constrain the haze flux well, ranging from 0 to 1.63 mJy, with corresponding values of Charon's diurnal thermal inertia from 5 to 23 MKS.

3. The lightcurve shape can provide useful constraints on Pluto's surface properties. Our best-fit model indicates a diurnal thermal inertia of 8--10 MKS on Pluto's surface and a low emissivity of CH$_4$ ice consistent with the small grain case.

4. JWST mid-infrared observations can separate Pluto from Charon and break the degeneracy of emission between Pluto's haze and Charon's surface. We predict that the rotational lightcurve of Pluto at 18 $\mu$m observed by JWST in 2022-2023 will be relatively flat, which could directly assess the existence of haze thermal emission as Pluto's surface emission is only around 0.1 mJy assuming the same spectral emissivity at 18 $\mu$m as at 24 $\mu$m.

5. The lightcurve variations of both Pluto and Charon in the JWST era range from tens of $\mu$Jy at 18 $\mu$m to hundreds of $\mu$Jy at 25.5 $\mu$m. Their variations can be characterized with JWST to constrain the thermal inertia and ice albedo distributions on both bodies.

Previous atmospheric models that explained the cold temperature of Pluto predict a wide range of haze emission at 24 $\mu$m. The haze flux is approximately 2.48 mJy if its particles are composed of Titan-like tholins. However, if the particles are composed of brighter ice particles, the flux could decrease by more than an order of magnitude outside the ice absorption band. The current observations cannot distinguish between the two scenarios, and thus cannot constrain the cooling rate in Pluto's atmosphere. At 18 $\mu$m, the haze flux is 1.5 (1.2) mJy for Titan-like tholins in 2004 (2023). Outside the ice absorption band, the flux could be smaller than 0.1 mJy for brighter ice particles. It is thus possible that JWST observations could help to distinguish between the two scenarios and provide more information about the cooling rate in Pluto's atmosphere. Specifically, the mid-IR spectra from JWST should be useful in identifying the solid-state features of ice particles, thereby aiding the differentiation between the two potential scenarios.

It is important to note the assumptions made in this study. First, we assume that Pluto's surface map has not changed significantly in the past two decades when calculating the lightcurves of Spitzer in 2004 and Herschel in 2012 (Figure \ref{fig3}), and JWST in 2023 (Figure \ref{fig5}), based on the derived map from New Horizons in 2015 (Figure \ref{fig1}). Although a time-invariant map in our model can explain all observations, Pluto's surface is expected to change over time. As the surface maps of Pluto and Charon are crucial inputs when modeling rotational emission lightcurves, a future study should include a seasonal change of the surface ice and albedo distribution. That would require a combined analysis of the New Horizons data revealing the limb-darkening information, multi-decadal ground-based observations of Pluto and Charon \citep{2003Icar..162..171B, 2015ApJ...804L...6B}, and an atmosphere-surface evolution model \citep{2015Icar..254..306T, 2017Icar..287...54F, 2017Icar..284..443Y, 2020JGRE..12506120B}.

Second, emissivity and thermal inertia are also important factors in determining surface temperature besides albedo. Due to limited information, we follow the approach of \citet{2011Icar..214..701L} and divide Pluto's surface into three component units based on the albedo map and assume the same emissivity within each unit. However, recent principal component analysis using the LEISA observations has revealed several more ice units \citep{2017Icar..287..229S, 2021Icar..35613833G, 2023PSJ.....4...15E}. Different mixtures of ice could have varying porosity and grain size, which would affect emissivity \citep{1991IJRS...12.2299L, 1996P&SS...44..945S}. Additionally, we assume the same thermal inertia for the entire globe, but it could vary with location and depth due to changes in composition. Future work with better data may relax these assumptions to retrieve the thermal inertia and emissivity of individual geological units and connect them with the ice compositions from spectroscopic measurements.

Finally, our model only accounts for the diurnal variation of surface and subsurface temperatures, as we focus on rotational lightcurves within a day. We assume that the heat flux from the deepest substrate in our model is zero, which is not entirely accurate due to the large thermal inertia in the deep layers of Pluto on a seasonal timescale. Although this assumption seems working well for the infrared data, seasonal changes have been observed in radio observations to probe the deep subsurface \citep[e.g.,][]{2011epsc.conf..271G, 2017DPS....4910202B, 2019Icar..322..192B}. Future work should consider both diurnal and seasonal changes like \citet{2016A&A...588A...2L}, and combine both infrared and radio data to finally understand the short-term and long-term evolution of the surface, subsurface, and atmospheric haze on Pluto.

\section{Acknowledgments}

This work is supported by NASA Solar System Workings Grant 80NSSC19K0791. We thank the reviewer for their valuable and constructive comments. We thank Dr. Bonnie Buratti for the discussions about albedo data from New Horizons. Retrievals are performed on lux supercomputer at UC Santa Cruz, funded by NSF MRI grant AST 1828315.


\appendix

\renewcommand{\thefigure}{A\arabic{figure}}
\setcounter{figure}{0}

\section{Validation of the self-rotating thermophysical model against an orbit-rotating model}
\label{sec:A1}

We compare our self-rotating thermophysical model used in the retrieval to that with the orbital revolution. Our simplified model includes only the shallow subsurface, the planetary self-rotation, and assuming zero flux from the deep subsurface. The orbit-rotating model, on the other hand, includes both orbital revolution and self-rotation of the planetary body. Besides the shallow subsurface, there is another deep subsurface where diurnal changes of incoming solar flux could hardly have an impact, as described in Section \ref{sec:2.2}.
Even though such simplification was suggested by \citet{1989Icar...78..337S}, a detailed comparison to the orbit-rotating model has not been reported. 

\begin{figure}[!htb]
\centering
\includegraphics[width=18cm]{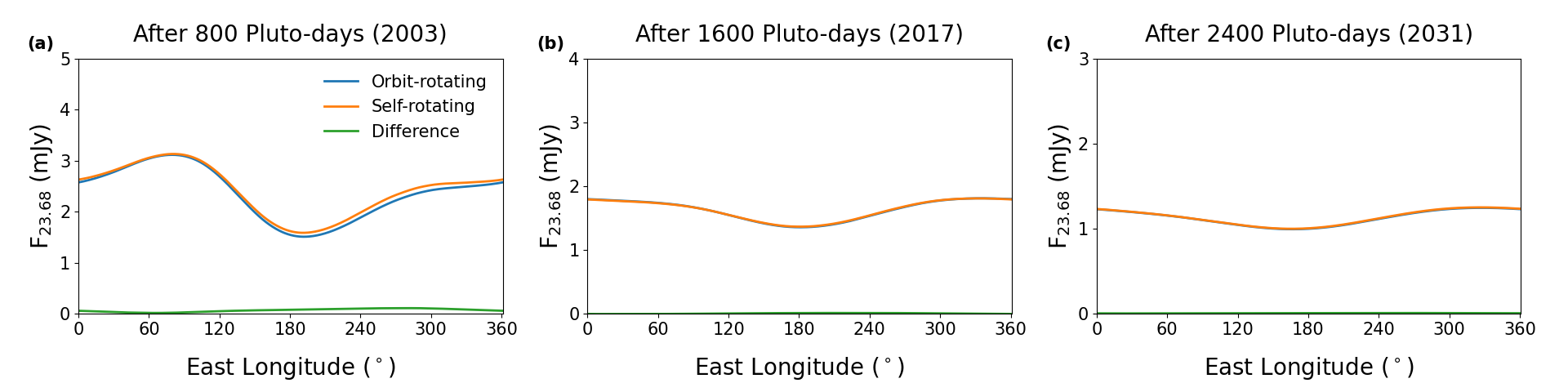}
\caption{Comparison of Pluto's rotational lightcurves from the orbit-rotating model (blue) and self-rotating model (orange) at 24 $\mu$m. Their difference (green) is less than 5\% of the total emission flux in 2003 and smaller than 1\% in 2017 and 2031.}
\label{fig7}
\end{figure}

In this validation, we chose an idealized setup. The input parameters of the orbit-rotating model with two layers are the same as described in Section \ref{sec:2.2}: $\Gamma_{day}$ = 20 MKS till the depth of 0.07 m and additional $\Gamma_{year}$ = 800 MKS underneath till 383.5 m. Other parameters of ice properties are the same as the ``free retrieval" scenario. 
We assume an orbit similar to Pluto's, which has an obliquity of 119.6$^\circ$, an eccentricity of 0.2444, and a semi-major axis of 39.48 AU.
We first run the two-layer orbit-rotating model using a coarse time step of 1 Pluto-day (6.4 Earth-days) for several orbits so that the deep layer temperatures reach the steady state when the deep temperatures at the same orbital position are consistent in two successive orbits. Running three orbital periods seems to be enough for such a long orbital period of 14179 self-rotating days from our test. Then we select a starting point at the orbit as the initial condition (e.g., perihelion in 1989), and run simulations with both orbital motion and self-rotation, in a finer time step of 1/1800 Pluto-day. Here we choose 800, 1600, and 2400 Pluto-days after perihelion as the final positions on Pluto's orbit for calculating lightcurves (blue in Figure \ref{fig7}). We also run our simplified one-layer model only with self-rotation at those final positions. After ensuring the diurnal temperature variation reaches a steady state, the self-rotating lightcurves (orange) are compared to orbit-rotating ones.
The difference between these two models (green) is smaller than 5\% in 2003 and less than 1\% in 2017 and 2031. As our retrieval requires a fast forward model, this difference is acceptable, trading off the accuracy and the efficiency.

\end{CJK*}
\end{document}